\documentstyle[11pt,epsfig]{article}
\title{\bf Noncommutativity Effects in FRW Scalar Field Cosmology}
\author{{\small Behrooz Malekolkalami}\footnote{b\_malekolkalami@sbu.ac.ir}\ \ {\small and
         Mehrdad Farhoudi}\footnote{m-farhoudi@sbu.ac.ir}\\
        {\small Department of Physics, Shahid Beheshti University, G.C.,}\\
        {\small Evin, Tehran 19839, Iran}}
\begin{document}
\date{\small May 16, 2009}
\maketitle %\baselineskip 24pt
\begin{abstract}
We study effects of noncommutativity on the phase space generated
by a non--minimal scalar field which is conformally coupled to the
background curvature in an isotropic and homogeneous FRW
cosmology. These effects are considered in two cases, when the
potential of scalar field has zero and nonzero constant values.
The investigation is carried out by means of a comparative
detailed analysis of mathematical features of the evolution of
universe and the most probable universe wave functions in
classically commutative and noncommutative frames and quantum
counterparts. The influence of noncommutativity is explored by the
two noncommutative parameters of space and momentum sectors with a
relative focus on the role of the noncommutative parameter of
momentum sector. The solutions are presented with some of their
numerical diagrams, in the commutative and noncommutative
scenarios, and their properties are compared. We find that impose
of noncommutativity in the momentum sector causes more ability in
tuning time solutions of variables in classical level, and has
more probable states of universe in quantum level. We also
demonstrate that special solutions in classical and allowed wave
functions in quantum models impose bounds on the values of
noncommutative parameters.
\end{abstract}
\bigskip
\section{Introduction}
\indent

Scalar fields are an integral part of modern models in particle
physics~\cite{3}, and recently play very important roles in
cosmology and have become a powerful tool to build cosmological
models as well. They have key role in some of these models as
current models of early cosmological inflation~\cite{1}, or, in
the viability of scalar field models as favorite candidates for
dark matter~\cite{2}. Scalar field cosmological models have
extensively been studied in the literatures, see, e.g.,
Refs.~\cite{4} and references therein. In the simplest
interactions, a scalar field is coupled to gravity. In many
cosmological models, scalar fields present degrees of freedom and
appear as dynamical variables of corresponding phase space, where
this point can be regarded as relevance of noncommutativity in
these models.

The proposal of noncommutativity concept between space--time
coordinates was introduced first by Snyder~\cite{5}, and about
twenty years ago, a mathematical theory, nowadays known as
noncommutative geometry (\textbf{NCG}), has begun to take
shape~\cite{6} based on this concept. In the last decade, study
and investigation of physical theories in the noncommutative ({\bf
NC}) frame, like string and M--theory~\cite{7,8}, has caused a
renewed interest on noncommutativity in the classical and quantum
fields. In particular, a novel interest has been developed in
considering the NC classical and quantum cosmology. In these
studies, the influence of noncommutativity has been explored by
the formulation of a version of NC cosmology in which a
deformation of minisuperspace~\cite{9}--\cite{11} or,  of phase
space~\cite{12} is required instead of space--time deformation.
From qualitative point of view, noncommutativity in the
configuration space leads to general effects, however, a
non--trivial noncommutativity in momentum sector introduces
distinct effects in what concern with the behavior of dynamical
variables.

Our purpose in this work is to build a NC scenario for the
Friedmann--Robertson--Walker~({\bf FRW}) cosmology including
matter field via a deformation achieved by the Moyal
product~\cite{7} in classical and quantum level. We introduce
effects of noncommutativity by two parameters, namely $\theta$ and
$\beta$, which are the NC parameters corresponding to space and
momentum sectors, respectively. Then, we will show that impose of
noncommutativity in the momentum sector causes more ability in
tuning time solutions of variables in classical level, and has
more probable states of universe in quantum level.

The work is organized as follows. In Section 2, we specify a model
and inspect it in the classical version within the commutative and
NC frames. Section 3 considers the quantum version of this model
by investigating universe wave functions and compares their
properties in the commutative and NC frames. A brief conclusion is
presented in the last section.

\section{The Classical Model}
\indent

We consider a classical model consisted of a cosmological system
that is presented by a four--dimensional action with a
non--minimally coupled scalar field  to gravity in a FRW universe.
To specify the NC effects of the model, we first treat the
commutative version, and then the NC one in the following
subsections.

\subsection{Commutative Phase Space}
\indent

A general action for a non--minimally coupled scalar field can be
described by
\begin{equation}\label{B1}
{\cal
A}=\int\sqrt{-g}\left[f(\phi)R-\frac{1}{2}g^{\mu\nu}\phi,_\mu
\phi,_\nu-V(\phi)\right]d^4x\, ,
\end{equation}
where $g$ is the determinant of the metric $g_{\mu\nu}$, $R$ is the
Ricci scalar, $V(\phi)$ and $f(\phi)$ are potential and coupling
functions of the scalar field, respectively. We assume a homogeneous
scalar field, that is $\phi=\phi(t)$, and the following FRW metric
of the minisuperspace
\begin{equation}\label{B2}
ds^2=-N^2(t)dt^2+a^2(t)\left(\frac{dr^2}{1-kr^2}+r^2d\Omega^2\right)\,
,
\end{equation}
where $N(t)$ is a lapse function, $a(t)$ is a scale factor and $k$
specifies geometry of the universe. Substituting the metric
(\ref{B2}) in action (\ref{B1}), one obtains the Lagrangian
density
\begin{equation}\label{B3}
{\cal L}
=6af\Bigl(kN-\frac{\dot{a}^2}{N}\Bigr)-6\frac{a^2\dot{a}\dot{f}}{N}+a^3N\Bigl(\frac{\dot{\phi}^2}{2N^2}-V\Bigr)\,
,
\end{equation}
where total time derivative terms have been neglected.

We restrict our considerations to the case of conformally coupled
scalar field model~\cite{11,barros98}. The general reason for
selecting such a scalar field is that it allows exact solutions in
simple cases, as those discussed along this work, and it is rich
enough to be useful as a probe for the significant modifications
that NCG introduces in classical and quantum cosmologies. That is,
we set $f(\phi)=1/(2\kappa)-\xi\phi^2/2$, where $\kappa=8\pi
G/c^4$ and $\xi$  is the non--minimal coupling parameter that
represents a direct coupling between the scalar field and
curvature, and has an arbitrary value. Obviously, the case $\xi=0$
is the minimally coupling situation, however, as mentioned, we
consider the conformal coupling case, i.e. $\xi=1/6$, and employ
the unites $\hbar=1=c$ and $\kappa=3$.

Based on these assumptions and rescaling the scalar field as
\begin{center}
$\chi=a\phi/\sqrt{2}$\,
\end{center}
the Lagrangian (\ref{B3}) reads
\begin{equation}\label{B4}
{\cal
L}=kNa-\frac{a\dot{a}^2}{N}+\frac{a\dot{\chi}^2}{N}-\frac{kN\chi^2}{a}-a^3NV\,
.
 \end{equation}
Thus, the corresponding Hamiltonian is
\begin{equation}\label{B5,1}
{\cal
H}=N\left(-\frac{p^2_a}{4a}+\frac{p^2_\chi}{4a}-ka+\frac{k\chi^2}{a}+a^3V\right),
\end{equation}
where  $p_a$ and $p_\chi$  are the canonical conjugate momenta.
For the conformal time gauge selection, namely $N=a$, one gets
\begin{equation}\label{B5,2}
{\cal H}=-\frac{p^2_a}{4}+\frac{p^2_\chi}{4}-ka^2+k\chi^2+a^4V\, .
\end{equation}
Then, the Hamilton equations are
\begin{eqnarray}\label{B6}
&&\dot{a}=\{a,{\cal H}\}=-\frac{1}{2}p_a\, , \cr
&&\dot{p_a}=\{p_a,{\cal H}\}=2ka-4a^3V+\sqrt{2}\chi a^2V'\, , \cr
&&\dot{\chi}=\{\chi,{\cal H}\}=\frac{1}{2}p_\chi\, ,\cr &&
\dot{p_\chi}=\{p_\chi,{\cal H}\}=-2k\chi-\sqrt{2}a^3V'\, ,
\end{eqnarray}
where the prime denotes derivative with respect to $\phi$.

In this work, in order to proceed further, we simply  treat two
special cases for the potential function, namely when there is no
potential and when there is a non--zero constant value potential,
$V=V_o$.

In free potential case, solutions to equations (\ref{B6}),
corresponding to the values of index curvature, with the
Hamiltonian constraint, ${\cal H} \approx 0$, are as follows
\begin{eqnarray}\label{C14}
k=1 :\left\{
\begin{array}{lll}
a(t)= A_1\cos t+A_2\sin t\hspace{.7cm}{\rm and}
\hspace{.7cm}\chi(t)=B_1\cos t+B_2\sin t\, ,
\\
\\
{\rm with\hspace{.2cm} constraint\!:}\hspace{.7cm}
A_1^2+A_2^2=B_1^2+B_2^2\, ,
\end{array}
\right.
\end{eqnarray}
\begin{eqnarray}\label{C141}
k=-1 :\left\{
\begin{array}{lll}
a(t)= A_3e^t+A_4e^{-t}\hspace{.7cm} {\rm and}
\hspace{.7cm}\chi(t)=B_3e^t+B_4e^{-t}\, ,
\\
\\
{\rm with\hspace{.2cm} constraint\!:}\hspace{.7cm} A_3A_4=B_3B_4\,
,
\end{array}
\right.
\end{eqnarray}
\begin{eqnarray}\label{C142}
k=0 :\left\{
\begin{array}{lll}
a(t)=A_5t+A_6\hspace{.7cm} {\rm and}
\hspace{.7cm}\chi(t)=B_5t+B_6\, ,
\\
\\
{\rm with\hspace{.2cm} constraint\!:}\hspace{.7cm} A_5^2=B_5^2\, ,
\end{array}
\right.
\end{eqnarray}
where $A_i$'s and $B_i$'s are constants of integration.

We will compare these solutions with their NC analogues in the next
section, where we will also discuss the case of non--zero constant
potential along with its NC correspondent.

\subsection{Noncommutative Phase Space}
\indent

Noncommutativity in classical physics is described by the Moyal
product law (shown by the $\ast$ notation in below) between two
arbitrary functions of phase space variables, namely
$\zeta^a=(x^i,p^j)$ for $i=1, \cdots, l$ and $j=l+1, \cdots, 2l$,
as~\cite{7}
\begin{equation}\label{A1}
(f\ast
g)(\zeta)=\exp\left[\frac{1}{2}\alpha^{ab}\partial_a^{(1)}\partial_b^{(2)}\right]f(\zeta_1)g(\zeta_2)
{\biggr|}_{\zeta_1=\zeta_2=\zeta}\, ,
\end{equation}
such that
\begin{equation}\label{A2}
(\alpha_{ab})=\left(%
\begin{array}{cc}
\theta_{ij} & \delta_{ij}+\sigma_{ij} \\-\delta_{ij}-\sigma_{ij}& \beta_{ij} \\
\end{array}
\right),
\end{equation}
where $a, b=1, 2, \cdots, 2l$, $\theta_{ij}$ and $\beta_{ij}$ are
assumed to be elements of real and antisymmetric matrices,
$\sigma_{ij}$ is a symmetric matrix (which can be written as a
combination of $\theta_{ij}$ and $\beta_{ij}$), and dimension of
the classical phase space is $2l$. The deformed or modified
Poisson brackets are defined as
\begin{equation}\label{A3}
\{f,g\}_\alpha=f\ast g-g\ast f\, ,
\end{equation}
and also the modified Poisson brackets of variables are
\begin{equation}\label{A4}
\{x_i,x_j\}_\alpha=\theta_{ij},\hspace{.5cm}\{x_i,p_j\}_\alpha=\delta_{ij}+\sigma_{ij}
\hspace{.5cm}{\rm and}\hspace{.5cm}\{p_i,p_j\}_\alpha=
\beta_{ij}\, .
\end{equation}

The simplest way to study physical theories within the NCG is
replacement of the Moyal product with the ordinary multiplication.
Actually, where variables of classical phase space obey the usual
Poisson brackets, i.e. $\{x_i,x_j\}=0=\{p_i,p_j\}$ and
$\{x_i,p_j\}=\delta_{ij}$, one can consider the following suitable
linear, non--canonical, transformation
\begin{equation}\label{A5}
x'_i=x_i-\frac{1}{2}\theta_{ij}p^j\hspace{.5cm}{\rm
and}\hspace{.5cm}p'_i=p_i+\frac{1}{2}\beta_{ij}x^j\, ,
\end{equation}
with constants $\theta_{ij}$ and $\beta_{ij}$, for which the usual
Poisson brackets of the primed variables yield
\begin{equation}\label{A6}
\{x'_i,x'_j\}=\theta_{ij},\hspace{.5cm}\{x'_i,p'_j\}=\delta_{ij}+\sigma_{ij}\hspace{.45cm}{\rm
and}\hspace{.45cm} \{p'_i,p'_j\}=\beta_{ij}\, ,
\end{equation}
where $\sigma_{ij}=-\theta_{k(i}\beta_{j)l}\delta^{kl}/4$. Hence,
relations (\ref{A6}) give the same results as relations
(\ref{A4}). That is, one can still employ the commutative
variables instead of the NC ones, where it is more convenient to
work with the Poisson brackets (\ref{A6}) than the modified
Poisson brackets (\ref{A4}). The transformation (\ref{A5}) can be
considered as an extension of the classical mechanics to the NC
classical mechanics. In geometrical language, the usual Poisson
brackets are mapped to the modified Poisson brackets through
transformation (\ref{A5}) and it should be noted that the two sets
of deformed and ordinary Poisson brackets represented by relations
(\ref{A4}) and (\ref{A6}) must be considered as distinct
relations. Although, for a compatible extension, the
transformation (\ref{A5}) must have an inverse and this imposes
some conditions on the NC parameters, which we will specify for
our model in below.

Furthermore, let ${\cal H}={\cal H}(x_i,p_i)$ be the Hamiltonian
of a system in the commutative case, we shift the canonical
variables through (\ref{A5}) and assume that the functional form
of the Hamiltonian in the NC case is still the same as the
commutative one, i.e.
\begin{equation}\label{C}
{\cal H}_{\rm nc}\equiv{\cal H}(x'_i,p'_i)={\cal
H}\left(x_i-\frac{1}{2}\theta_{ij}p^j,\,
p_i+\frac{1}{2}\beta_{ij}x^j\right).
\end{equation}
This function is defined on the commutative space and therefore,
the equations of motion for unprimed variables are obviously
$\dot{x}^i=\partial{\cal H}_{\rm nc}/\partial p_i$\ and
$\dot{p}^i=-\partial{\cal H}_{\rm nc}/\partial x_i$. Evidently,
the effects due to the noncommutativity arise by terms including
the parameters $\theta_{ij}$ and $\beta_{ij}$.

Let us now proceed to study the behavior of the model in a phase
space with deformed Poisson brackets (\ref{A6}) such that the
minisuperspace variables do~not commute with each other. For our
model, the two gravitational degrees of freedom are the
minisuperspace coordinates $(a,\chi)$. The corresponding phase
space variables are $(x^1,x^2,p^1,p^2)=(a,\chi,p_a,p_\chi)$, where
$p_a$ and $p_\chi$ are the linear minisuperspace momenta. We
assume the NC parameters
\begin{equation}\label{ourassumpyion}
\theta^{12}\equiv\theta\geq0\quad {\rm and}\quad
\beta^{12}\equiv4\beta\geq0\, ,
\end{equation}
hence $\sigma^{12}=\theta\beta$, where $\theta$ and $\beta$ are
constants. Thus, after making the transformations
\begin{equation}\label{C3}
a\rightarrow
a-\frac{\theta}{2}p_\chi\hspace{0.1cm},\hspace{0.5cm}p_a\rightarrow
p_a+2\beta\chi \hspace{0.1cm},\hspace{0.5cm}
\chi\rightarrow\chi+\frac{\theta}{2}p_a\hspace{0.4cm} {\rm and}
\hspace{0.4cm}p_\chi \rightarrow p_\chi-2\beta a,
\end{equation}
in (\ref{B5,2}), the NC Hamiltonian will be
\begin{equation}\label{C4}
{\cal H}_{\rm
nc}=-\frac{1-k\theta^2}{4}(p_a^2-p_\chi^2)+(k\theta-\beta)(\chi
p_a+ap_\chi)+(\beta^2-k)(a^2-\chi^2)+(a-\frac{1}{2}\theta
p_\chi)^4\tilde{V}\, ,
\end{equation}
where $\tilde{V}=V(\tilde{\phi})$ is a modified potential with
$\tilde{\phi}=\sqrt{2}(\chi+\theta p_a/2)/(a-\theta p_\chi/2)$.
Besides, the inverse transformation of (\ref{C3}) exists when its
determinant is~not zero, that is when $\theta\beta\neq1$. Now, the
equations of motion for the NC Hamiltonian are
\begin{eqnarray}\label{C5}
&&\dot{a}=-\frac{1}{2}(1-k\theta^2)p_a+(k\theta-\beta)\chi+\frac{1}{\sqrt{2}}
\theta(a-\frac{1}{2}\theta p_\chi)^3\tilde{V}'\, ,\cr
&&\dot{p_a}=2(k-\beta^2)a+(\beta-k\theta)p_\chi-4(a-\frac{1}{2}\theta
p_\chi)^3V+\sqrt{2}(\chi+\frac{1}{2}\theta
p_a)(a-\frac{1}{2}\theta p_\chi)^2\tilde{V}'\, ,\cr
&&\dot{\chi}=(k\theta-\beta)a+\frac{1}{2}(1-k\theta^2)p_\chi-2\theta(a-\frac{1}{2}\theta
p_\chi)^3V+\frac{1}{\sqrt{2}}\theta(\chi+\frac{1}{2}\theta
p_a)(a-\frac{1}{2}\theta p_\chi)^2\tilde{V}'\, , \cr
&&\dot{p_\chi}=(\beta-k\theta)p_a+2(\beta^2-k)\chi-\sqrt{2}(a-\frac{1}{2}\theta
p_\chi)^3\tilde{V}'\, ,
\end{eqnarray}
where $\tilde{V}'=dV/d\tilde{\phi}$. Comparing equations
(\ref{C5}) with (\ref{B6}) shows that, in general, equations of
motion are coupled in the NC case and for $\theta=0=\beta$,
equations (\ref{C5}) reduce to (\ref{B6}) as expected.

In free potential case, equations (\ref{C5}) read
\begin{eqnarray}\label{C9}
&&\dot{a}=-\frac{1}{2}(1-k\theta^2)p_a+(k\theta-\beta)\chi\hspace{.1cm}
, \cr &&\dot{p_a}=2(k-\beta^2)a+(\beta-k\theta) p_\chi
\hspace{.1cm}, \cr
&&\dot{\chi}=(k\theta-\beta)a+\frac{1}{2}(1-k\theta^2)p_\chi\hspace{.1cm}
, \cr &&\dot{p_\chi}=(\beta-k\theta)p_a+2(\beta^2-k)\chi\, .
\end{eqnarray}
By eliminating momenta variables, one gets
\begin{equation}\label{C13}
\ddot{a}=-m^2a+2(k\theta-\beta )\dot{\chi}\hspace{1cm}{\rm and}
\hspace{1cm}\ddot{\chi}=-m^2\chi+2(k\theta-\beta)\dot{a}\hspace{.1cm},
\end{equation}
with the Hamiltonian constraint
\begin{center}
$(\beta- 1)(a^2-\chi^2)={\rm constant}$  \hspace{1.9cm}{\rm if}
\hspace{.4cm} $1-k\theta^2=0$\ \rlap,\footnote{Note that, the
condition $1-k\theta^2=0$ is possible only when $k=1$ and hence
$\theta=1$.}
\end{center}
or
\begin{equation}
m^2(a^2-\chi^2)-(\dot{a}^2-\dot{\chi}^2)=0\hspace{1.7cm}{\rm if}
\hspace{.5cm}1-k\theta^2\neq0\, ,
\end{equation}
where $m^2\equiv k(1-\beta\theta)^2$. It is noticeable that when
$k=0$, the NC parameter $\theta$ is removed from equations
(\ref{C13}), that is, in the spatially flat FRW universe, the
motion equations are affected only by the NC parameter $\beta$.

If one takes $\Delta\equiv(1-k\theta^2)(\beta^2-k)$, real
solutions of equations (\ref{C13}) can be written as
\begin{equation}\label{C15}
 \Delta>0:\!\cases{
 a(t)=e^{(k\theta-\beta)t}(A_1\sinh\sqrt{\Delta}t+
 B_1\cosh\sqrt{\Delta}t)
 +e^{-(k\theta-\beta)t}(C_1\sinh\sqrt{\Delta}t+
 D_1\cosh\sqrt{\Delta}t)\cr
 \cr
 \chi(t)=e^{(k\theta-\beta)t}(A_1\sinh\sqrt{\Delta}t+
 B_1\cosh\sqrt{\Delta}t)
 -e^{-(k\theta-\beta)t}(C_1\sinh\sqrt{\Delta}t+
 D_1\cosh\sqrt{\Delta}t)\cr}
\end{equation}
and
\begin{equation}\label{C16}
 \Delta<0:\!\cases{ a(t)=
 e^{(k\theta-\beta)t}(A_2\sin\sqrt{-\Delta}t+
 B_2\cos\sqrt{-\Delta}t)+e^{-(k\theta-\beta)t}(C_2\sin\sqrt{-\Delta}t+
 D_2\cos\sqrt{-\Delta}t)\cr
 \cr
 \chi(t)=e^{(k\theta-\beta)t}(A_2\sin\sqrt{-\Delta}t+
 B_2\cos\sqrt{-\Delta}t)-e^{-(k\theta-\beta)t}(C_2\sin\sqrt{-\Delta}t+
 D_2\cos\sqrt{-\Delta}t),\cr}
\end{equation}
where $A_i$'s, $B_i$'s, $C_i$'s and $D_i$'s are constants of
integration.

The quantity $\Delta$ is always positive when $k=0, -1$, hence, in
general, solutions (\ref{C15}) are in the form of
\begin{equation}\label{C17}
\sum^2_{j=1}\left(b\, e^{\delta_j t}+c\, e^{-\delta_j t}\right)\,
,
\end{equation}
with $b$ and $c$ as new constants, and
\begin{center}
$\delta_1\equiv k\theta-\beta+\sqrt{\Delta}$\hspace{4mm} and
\hspace{4mm}$\delta_2\equiv k\theta-\beta-\sqrt{\Delta}$\, .
\end{center}
Comparing solutions (\ref{C17}) with its commutative analogues
(\ref{C141}) and (\ref{C142}) shows that for $ k=-1$, solutions
are still hyperbolic, but with an extra coefficient in the
exponent that depends on the NC parameters and gives enough room
for better adjustments. In the case $k=0$, one has $\delta_1=0$
and $\delta_2=-2\beta$, and hence, the time dependence of
solutions have been modified from linear in commutative case to
hyperbolic in the NC case, where the former solution is~not
suitable for an accelerating universe, though the latter one is
capable to be adjusted with the observed accelerated expansion.
Also, in the late time, solution (\ref{C17}) describes a de Sitter
universe for which its cosmological constant is written in terms
of the NC parameter as $\Lambda/3=4\beta^2$.

For $k=1$ geometry, one has $\Delta=(1-\theta^2)(\beta^2-1)$,
which can be positive or negative or zero, depends on the
different choices of the NC parameters. That is, when
$(\theta>1,\hspace{2mm}\beta<1)$ or
$(\theta<1,\hspace{2mm}\beta>1)$, $\Delta$ is positive. Hence, we
have solutions (\ref{C15}), however, in this case the NC
parameters have upper and lower bounds. When $\Delta<0$, existence
of oscillating solutions is provided when both NC parameters
simultaneously have a lower bound, namely $\theta$ \textsl{and}
$\beta>1$, or an upper bound, namely $\theta$ \textsl{and}
$\beta<1$. The commutative solutions (\ref{C14}) are oscillating
with the period of $2\pi$ and constant amplitudes, whereas the NC
solutions (\ref{C16}) have the period of $2\pi/\sqrt{-\Delta}$ and
varying amplitudes with factor $e^{(\theta-\beta)t}$. Note that,
the choice $\theta=\beta$, leads to a constant amplitude as the
commutative case, but with a period of $2\pi/|1-\theta^2|$. The
case $\Delta=0$ is possible when $\theta=1$ or when $\beta=1$,
where solutions are again hyperbolic. The case $k=1$ and
$\theta=1$, resembles the condition $1-k\theta^2=0$ with arbitrary
$\beta$, however, with the value of $\beta=1$ one gets trivial
constant solutions.

Now, we treat a non--zero positive constant potential $V=V_o$.
Since, in general, the motion equations are~not sufficiently
simple to be solved, we restrict ourself to the case $k=0$ and
$\beta=0$. Thus, after a little algebra, equations (\ref{C5}) give
\begin{equation}\label{C28}
\dot{a}^2=V_o(a-v_o\theta)^4+d\hspace{0.65cm} {\rm and}
\hspace{0.65cm}\chi(t)=\chi_o+v_ot-\theta \dot{a}\, ,
\end{equation}
where $\chi_o, d$ and $v_o\equiv p_\chi/2$ are constants and the
Hamiltonian constraint gives $d=v_o^2$.

By setting $\theta=0$, one obtains the commutative equations as
\begin{equation}\label{B12}
\dot{a}^2=V_oa^4+v_o^2\hspace{0.65cm} {\rm and}
\hspace{0.65cm}\chi(t)=\chi_o+v_ot\, .
\end{equation}
Solutions of $a(t)$ are in terms of the Jacobi elliptic functions,
i.e.
\begin{equation}\label{b12}
a(t)=\pm\, D\ {\bf\textsf{sn}}\left(i\sqrt{V_o}Dt, i\right)\,,
\end{equation}
with $D\equiv[v_o/(i\sqrt{V_o}\,)]^{1/2}$ and the initial
condition $a(0)=0$. These two solutions are periodic functions
with respect to the time, and for $V_0<0$, they qualitatively
behave as sine functions with amplitude $|D|$. For $V_0>0$,
solution (\ref{b12}) qualitatively behaves as a tangent function
and its diagram is plotted in Fig.~$1$ (left) for numerical values
$v_o=1$, $\chi_o=2$, $\theta=2$ and $V_o=1$. Obviously, physical
solution corresponds to the positive part of this diagram.

Equations (\ref{C28}) and (\ref{B12}) show that the effect due to
the noncommutativity is a shift in the scale factor with a
magnitude $|v_o\theta|$ that changes the zero point of the scale
factor. This effect can remove a negative scale factor when the
constant potential is negative. That is, in such a case, the
corresponding NC solution is $v_0\theta\pm D\
{\bf\textsf{sn}}\left(i\sqrt{V_o}Dt, i\right)$, where by choosing
$v_0\theta\geq |D|$ with $v_0>0$, then the resulted scale factor
will be positive. Clearly, plot of $a(t)$ in the NC case is the
same as the commutative one with a vertical transfer of magnitude.

The linear behavior of the scalar field in the commutative case is
altered by an additional term $-\theta\dot{a}$. Two plots of the
scalar field in the NC case are sketched in Fig.~$1$ (right) for
numerical values $v_o=1$, $\chi_o=2$ and $\theta=2$, one with
$V_0=-1$ (solid line) and one with $V_0=+1$ (dashed line). Note
that, in a positive potential, the scalar field goes to
$\pm\infty$ for very late time (when $t$ goes to infinity),
whereas, in a negative potential, it goes to $+\infty$
monotonically.

In terms of the cosmic time, $d\tau=adt$, the first equation in
(\ref{B12}) may have a solution, for $V_0>0$, as
\begin{equation}\label{B13}
a(\tau)\propto\Big[\sinh\left(2\sqrt{V_o}\tau\right)\Big]^{1/2}\,
.
\end{equation}
The behavior of such a scale factor in the late cosmic time, i.e.
$\tau\gg0$, is
\begin{equation}\label{B14}
a(t)\propto {1\over t_1-t}\ ,
\end{equation}
where $t_1$ is a constant such that if $\tau\rightarrow\infty$,
then $t\rightarrow t_1$. Hence, in this limit, the scalar field in
the NC case grows proportion to $\theta$ as $\theta(t_1-t)^{-2}$,
and deviation from the linear time dependence becomes very large,
as one can recognize it in Fig.~$1$ (left).

\section{The Quantum Model}
\indent

We proceed to quantize the cosmological model given by the action
(\ref{B1}) in the case of free potential, such that the canonical
quantization of the phase space leads to the Wheeler--DeWitt
(\textbf{WD}) equation, $\hat{{\cal H}}\Psi=0$, where $\hat{{\cal
H}}$ is the Hamiltonian operator and $\Psi$ is a wave function of
universe. For arguments about the quantization based on WD
equation in the FRW universe including matter fields, see, e.g.,
Refs.~\cite{13}. We employ the usual canonical transition from
classical to quantum mechanics via the generalized Dirac
quantization from the Poisson brackets to the quantum commutators,
i.e. $\{\}\rightarrow-i[\hspace{.1cm}]$. Then, as the classical
approach, we investigate the commutative and NC frames in the
following subsections.

\subsection{\bf Commutative Frame}
\indent

As usual, the operator form of Hamiltonian (\ref{B5,2}) can be
acquired by the replacements $ p_a\rightarrow -i\partial_a$ and
$p_\chi\rightarrow -i\partial_\chi$. Assuming a particular factor
ordering, the corresponding WD equation, for $V=0$, is
\begin{equation}\label{D1}
\left[\frac{\partial^2}{\partial a^2}-\frac{\partial^2}{\partial
\chi^2}+4k(\chi^2-a^2)\right]\Psi(a,\chi)=0\, .
\end{equation}
Considering the following change of variables
\begin{equation}\label{D2}
a=\rho\cosh\varphi\hspace{.75cm}{\rm and} \hspace{.75cm}
\chi=\rho\sinh\varphi\hspace{.1cm}\, ,
\end{equation}
equation (\ref{D1}) reads
\begin{equation}\label{D3}
\left(\frac{\partial^2}{\partial
\rho^2}+\frac{1}{\rho}\frac{\partial}{\partial
\rho}-\frac{1}{\rho^2}\frac{\partial^2}{\partial
\varphi^2}-4k\rho^2\right)\Psi(\rho,\varphi)=0\, .
\end{equation}

We assume solutions of equation (\ref{D3}) as a product {\it
ansatz}
\begin{equation}\label{D4}
\Psi(\rho,\varphi)=\psi(\rho)e^{2i\alpha\varphi}\, ,
\end{equation}
where $\alpha$ is a real constant. Substitution of (\ref{D4}) in
equation (\ref{D3}) leads to
\begin{equation}\label{D5}
\psi''+\frac{\psi'}{\rho}+4\Big(\frac{\alpha^2}{\rho^2}-k\rho^2\Big)\psi=0\,
,
\end{equation}
where the prime denotes derivative with respect to $\rho$. The
well--defined eigenfunctions of equation (\ref{D5}), considering
boundary conditions, can be written as
\begin{eqnarray}\label{D71}
\psi_\alpha(\rho)\propto\left\{
\begin{array}{lll}
J_{i\alpha}(-k\rho^2)\hspace{1.1cm}{\rm for}~~~~k=-1
\\
\\
\cos(2\alpha\ln\rho)\hspace{0.9cm}{\rm for}~~~~k=0
\\
\\
K_{i\alpha}(k\rho^2) \hspace{1.3cm}{\rm for}~~~~k=1\, ,
\end{array}
\right.
\end{eqnarray}
where the functions $J_\nu$ and $K_\nu$ are the first and modified
second kind of the Bessel functions, respectively. Hence, the wave
packet corresponding to (\ref{D71}) is
\begin{equation}\label{D72}
\Psi(\rho,
\varphi)=\int^{+\infty}_{-\infty}C_\alpha\psi_\alpha(\rho)e^{2i\alpha\varphi}d\alpha\,
,
\end{equation}
where $C_\alpha$ can be taken~\cite{9,10} to be a shifted Gaussian
weight function with constants $b$ and $c$, i.e.
$e^{-b(\alpha-c)^2}$.

Fig.~$2$ shows plots of the probability $|\Psi|^2$ for $k=0$ and
$k=1$ with values $b=1=c$, in the range of $0<\rho<10$ and
$-5<\varphi<5$. The corresponding plot for $k=-1$ is qualitatively
similar to the case $k=1$.

\subsection{\bf Noncommutative Frame}
\indent

We assume that, in a general NC quantum phase space, the
coordinate operators of the FRW minisuperspace and their
generalized momenta obey the star deformed Heisenberg algebra,
like the ones in noncommutative quantum mechanics, as~\cite{14}
\begin{equation}\label{A8}
[\hat{x}_i,
\hat{x}_j]_\alpha=i\theta_{ij},\hspace{.5cm}[\hat{x}_i,\hat{p}_j]_\alpha=i(\delta_{ij}+\sigma_{ij})
\hspace{.5cm}{\rm
and}\hspace{.5cm}[\hat{p}_i,\hat{p}_j]_\alpha=i\beta_{ij}\, .
\end{equation}
The notations and definitions are the same as in the NC classical
model. Now, the corresponding noncommutative WD equation can be
written by replacing operator product with the Moyal product,
namely
\begin{equation}\label{A91}
\hat{{\cal H}}_{\rm nc}(\hat{x}_i,\hat{p}_i)\ast\Psi(x,p)=0\, .
\end{equation}
It is well--known in noncommutative quantum mechanics~\cite{14}
that the original phase space and its symplectic structure can be
modified to reformulate relations in the commutative algebra when
the new variables, as in (\ref{A5}), are introduced. Hence, the
original equation (\ref{A91}) for these new variables
reads~\cite{15}
\begin{equation}\label{A92}
\hat{{\cal H}}_{\rm nc}\left(\hat{x}_i-\frac{1}{2}\theta_{ij}
\hat{p}^j,\hat{p}_i+\frac{1}{2}\beta_{ij}\hat{x}^j\right)\Psi(x,p)=0\,
.
\end{equation}
Therefore, the noncommutative WD equation corresponding to
relation (\ref{C4}), with $\theta\beta\neq1$ and for $V=0$, can be
written as
\begin{equation}\label{D9}
\left[(1-k\theta^2)(\partial^2_a-\partial^2_\chi)-
4i(k\theta-\beta)(\chi\partial_a+a\partial_\chi)+4(\beta^2-k)(a^2-\chi^2)\right]\Psi(a,\chi)=0\,
,
\end{equation}
that, with the change of variables (\ref{D2}), reads
\begin{equation}\label{D10}
\left[(1-k\theta^2)\Big(\frac{\partial^2}{\partial
\rho^2}+\frac{1}{\rho}\frac{\partial}{\partial
\rho}-\frac{1}{\rho^2}\frac{\partial^2}{\partial \varphi^2}\Big)-
4i(k\theta-\beta)\frac{\partial}{\partial
\varphi}-4(k-\beta^2)\rho^2\right]\Psi=0\, .
\end{equation}

If $1-k\theta^2\neq 0$, by the {\it ansatz} (\ref{D4}), equation
(\ref{D10}) reduces to
\begin{equation}\label{D11}
\psi''+\frac{\psi'}{\rho}+
4\left(\frac{\alpha^2}{\rho^2}-\epsilon\rho^2+2\gamma\alpha\right)\psi=0\,
,
\end{equation}
where $\epsilon\equiv (k-\beta^2)/(1-k\theta^2)$ and $\gamma\equiv
(k\theta-\beta)/(1-k\theta^2)$. Before discussing solutions of
(\ref{D11}), we enumerate a few features of this equation.
Firstly, it reduces to equation (\ref{D5}) when one sets
$\epsilon=k$ and $\gamma=0$, where a trivial case is
$\theta=0=\beta$, as expected. Secondly, it is interesting that,
in the two following cases where the noncommutativity is still
present, equation (\ref{D11}) again reduces to equation
(\ref{D5}). These two cases\footnote{The case $k=-1$ with
$\theta=-\beta$ is~not valid, for our
assumption~(\ref{ourassumpyion}).}
 are $k=1$ with $\theta=\beta\neq 1$, and $k=0$ with $\beta=0$ and arbitrary
$\theta$. In another words, in these cases, general form of the NC
wave functions are the same as the commutative
solutions~(\ref{D71}). Finally, in the case $k=0$, solutions to
equation (\ref{D11}) does~not depend on the $\theta$ parameter,
which is also a common feature in the classical model.

A particular solution of equation (\ref{D11}) can be written in
terms of the Whittaker functions, $W_{\mu\, ,\nu}$\ and $M_{\mu\,
,\nu}$, as
\begin{equation}\label{D12}
\psi_\alpha\left(\rho\right)=\rho^{-1}\left[A_{\alpha}M_{\mu\,
,\nu}\left(2\sqrt{\epsilon} \rho^{2}\right) +B_{\alpha}W_{\mu\,
,\nu}\left( 2\sqrt{\epsilon} \rho^{2}\right)\right]\, ,
\end{equation}
where $A_{\alpha}$ and $B_{\alpha}$ are superposition constants,
$\mu\equiv\alpha\gamma/\sqrt{\epsilon}$\ and $\nu\equiv i\alpha$.
The both Whittaker functions, even in classically forbidden
regions, are convergent when $\epsilon$ is
negative\rlap,\footnote{For properties of the Whittaker functions
and their indices, see, e.g., Refs.~\cite{16}.}
 that can obviously occur for $k=0$ and $k=-1$.
Also, $\epsilon$ can be negative in the case $k=1$ for special
values of the NC parameters, we will discuss these situations in
below.

In the case $k=0$ for which $\epsilon=-\beta^2$, one can write
solution (\ref{D12}) with one of the Whittaker functions, e.g.,
$M_{\mu\, ,\nu}$ and gets its corresponding wave packet as
\begin{equation}\label{D13}
\Psi(\rho,\varphi)=\rho^{-1}\int^\infty_{-\infty}e^{-b(\alpha-c)^2}
M_{i\alpha\, ,i\alpha}\left(2i\beta
\rho^{2}\right)e^{2i\alpha\varphi}d\alpha\, .
\end{equation}
Fig.~$3$ shows the probability $|\Psi|^2$ corresponding to
(\ref{D13}) for $\beta=1/10$ and the Gaussian constants $b=1=c$.
The comparison of this figure with the left diagram in Fig.~$2$
identifies that, although the most probable state of a commutative
universe is around $\varphi=0$, the most probable state related to
a NC universe is shifted for $\varphi$ less than zero. Also,
another difference in those two figures is the displacement of
peaks to greater values of $\rho$ in a NC universe with respect to
a commutative one. By consecutive diagrams, it can be shown that
this difference is also more manifested when $\beta$ gets smaller
values.

As mentioned before, the case $\epsilon<0$ can also be attained
when $k=1$, for which $\epsilon=(1-\beta^2)/(1-\theta^2)$. This
quantity is negative when one of the inequalities ($\theta<1$ and
$\beta>1$) or ($\theta>1$ and $\beta<1$) are held. The first index
of the Whittaker functions, for both of these inequalities, is
also a pure imaginary number. When $\theta\beta=1$, wave packets
resemble those of the case $k=0$, but this choice violates the
existence of the inverse transformation for (\ref{C3}). On the
other hand, plots of wave packets for $\theta\beta\neq1$, in
general, show that they are~not suitable for description of a real
universe. However, choosing special values, such as considering
the first inequality with $\theta=0$ and $\beta$ very large values
that $\beta^2\pm1\approx\beta^2$ or considering the second
inequality with $\beta=0$ and $\theta$ very large values that
$\theta^2\pm1\approx\theta^2$, one can get $\mu\approx i\alpha$
and hence again, construct well--behaved wave packets in a closed
universe, which their corresponding plots are qualitatively
similar to Fig.~3. For instance, Fig.~3 does also illustrate a
similar plot of probability for $\beta\approx 0$ and
$\theta\approx 10$ in this case, where the visible contrast
between such a figure and its commutative analogue, Fig.~$2$
(right), implies that the probability of expansion in a NC
universe is more than a commutative counterpart. Also, the most
probable state of a commutative universe is around $\varphi=0$ and
$\rho=0$, but the most probable state related to a NC universe is
shifted for $\varphi$ less than zero and $\rho$ greater than zero.
Besides, there is displacement of peaks to greater values of
$\rho$ in a NC universe with respect to a commutative one. This
effect is more manifested when $\theta$ gets greater values.
Hence, there are different possible universes (states) from which,
our present universe could be evolved and tunneled in the past
from one state to another\rlap.\footnote{Such a kind of arguments
can also be found in, e.g., Ref.~\cite{10}.}

The above discussion, in the latter paragraph, is completely
fulfilled when $k=-1$, except that there is~no bound on the NC
parameters, for $\epsilon$ is always negative in this case.

For when $\epsilon$ is positive, the Whittaker term of
$M_{\mu,\nu}$, in classically forbidden region, is divergent.
Besides, the both Whittaker terms, when $\epsilon>0$, are
proportional to $\exp (-\sqrt{\epsilon}\rho^2)$, and hence, are
quickly damped as $\rho$ grows.

We should mention that $\epsilon$ also vanishes for the case $k=1$
with $\beta=1$, which leads to $\gamma=-1/(1+\theta)$. In this
case, solution of equation (\ref{D11}) can be considered as
\begin{equation}\label{D141}
\psi_\alpha(\rho)\propto
J_{i\alpha}\left(2i\gamma\sqrt{\alpha}\rho\right)\, .
\end{equation}
This solution, as can easily be checked, is~not well defined for
constructing wave packets and hence, we do~not consider it.

The condition $1-k\theta^2=0$ is equivalent to $k=1$ and
$\theta=1$, for which equation (\ref{D10}) reduces to
\begin{equation}\label{D15}
\left[i(1-\beta)\frac{\partial}{\partial
\varphi}+(1-\beta^2)\rho^2\right]\Psi=0\, .
\end{equation}
Its solution is
\begin{equation}\label{D16}
\Psi=R(\rho)e^{i(1+ \beta)\varphi\rho^2}\, ,
\end{equation}
where $R(\rho)$ is any differentiable function. The probability of
this wave function is independent of the $\beta$ parameter.

\section{Conclusions}
\indent

In this work, we have carried out an investigation into the role
of NCG in cosmological scenario by introducing a NC deformation in
the algebra of phase space variables almost along the same lines
proposed in Ref.~\cite{11}. The space is generated by a conformal
scalar field when is non--minimally coupled to geometry whose
background is the FRW metric, in all three cases $k=-1, 0, 1$. The
noncommutativity has been introduced in the both, space and
momentum, sectors with more attention on the role of the NC
parameter $\beta$. The investigation has been carried out by means
of a comparative analysis of the evolution of universe and the
most probable universe wave functions in classically commutative
and NC frames and quantum counterparts.

In absence of the scalar field potential function, the classical
exact solutions have been obtained in both commutative and NC
frames. Contrary to the commutative case, the NC solutions can be
regulated with the aid of both NC parameters. Especially, in
spatially flat universes, the major player in tuning solutions is
the $\beta$ parameter, where it can be employed to adjust time
dependent solutions with the observational data, e.g., the
accelerating expansion rate. Interestingly, we have found that the
existence of particular solutions imposes bounds on the values of
NC parameters.

In the presence of a non--zero constant potential and absence of
$\beta$, the time dependence of the scalar field changes contrary
to the scale factor which is just shifted by a constant value
proportional to the $\theta$ parameter. This shift can remove
negative scale factors when negative constant potentials are used.
The scalar field deviation from linear time dependence of the
commutative case, for a positive constant potential, becomes more
apparent in the late cosmic time.

One expects that when noncommutativity effects are turned on in
the quantum scenario they should introduce significant
modifications. To see this, the corresponding quantum cosmology
has also been considered and the exact solutions, through the WD
equation in the commutative and NC frames, are obtained in order
to investigate differences in the solutions. As in the case of
classical model, the $\beta$ parameter is the only one that is
responsible for NC effects in spatially flat universes. Comparing
the numerical diagrams of the probability in commutative and NC
cases, especially in the case of positive and negative curvature
indexes, shows that expansion in a NC universe is more probable,
depending on the values of the NC parameters, than a commutative
counterpart. In the flat case, existence of universes with more
possible states invokes smaller values of $\beta$ in NC frames.
Also, existence of more probable states and transferring peaks of
probability to greater values of $\rho$, defined in (\ref{D2}), in
the NC frames with respect to the commutative ones are intensified
when $\beta$ gets smaller values. Also, the latter property can
qualitatively be apparent in the closed and open cases when, for
instance, $\beta=0$ with $\theta$ sufficiently large values, or
$\theta=0$ with $\beta$ sufficiently large values.
%%%%%%%%%%%%%%%%%%%%%%%%%%

%
\begin{figure}
\begin{tabular}{ccc}
\epsfig{figure=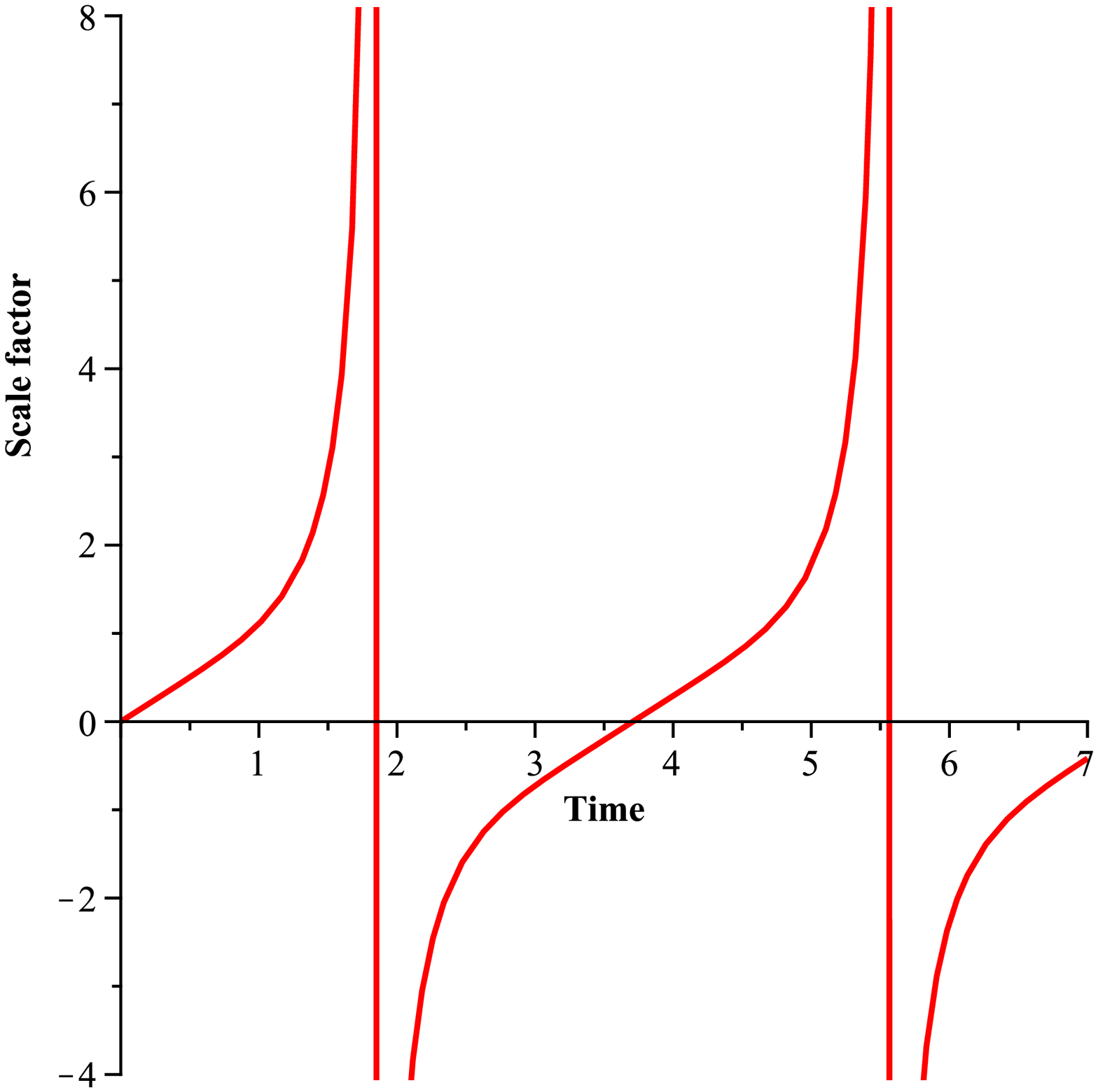,width=7cm}\hspace{1cm}
\epsfig{figure=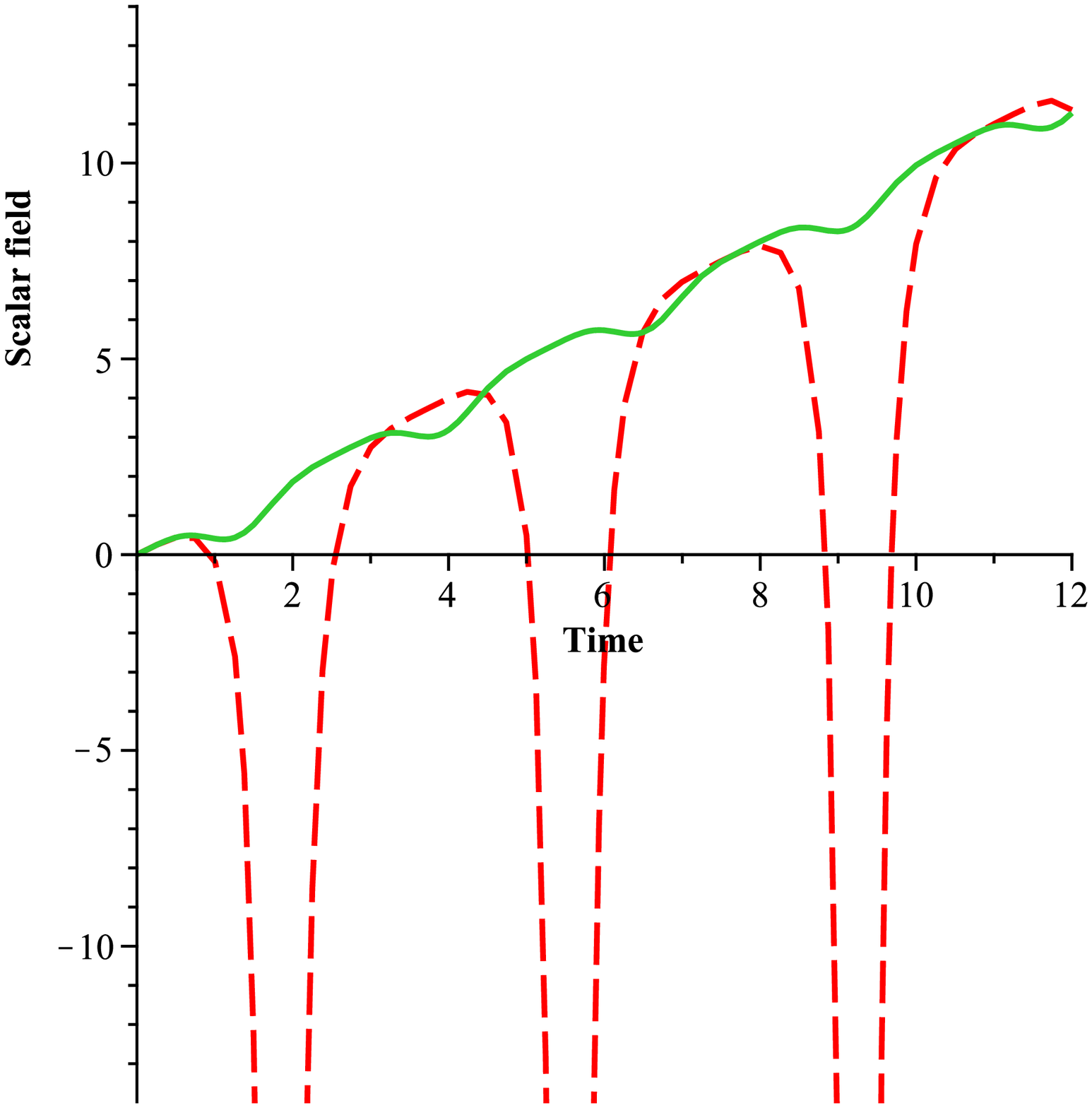,width=7cm}\vspace{1cm}
\end{tabular}

\hspace{0.5cm}{\footnotesize \textbf{Figure 1}: The scale factor
in the commutative case for $V_0=1$ (left) and the scalar field in
the NC case (right)

\hspace{0.5cm}for $V_0=-1$ (solid line) and $V_0=1$ (dashed line),
all plots with numerical values $v_o=1,\hspace{2mm}\chi_o=2$ and
$\theta=2$.}
\end{figure}
\begin{figure}
\begin{center}
\epsfig{figure=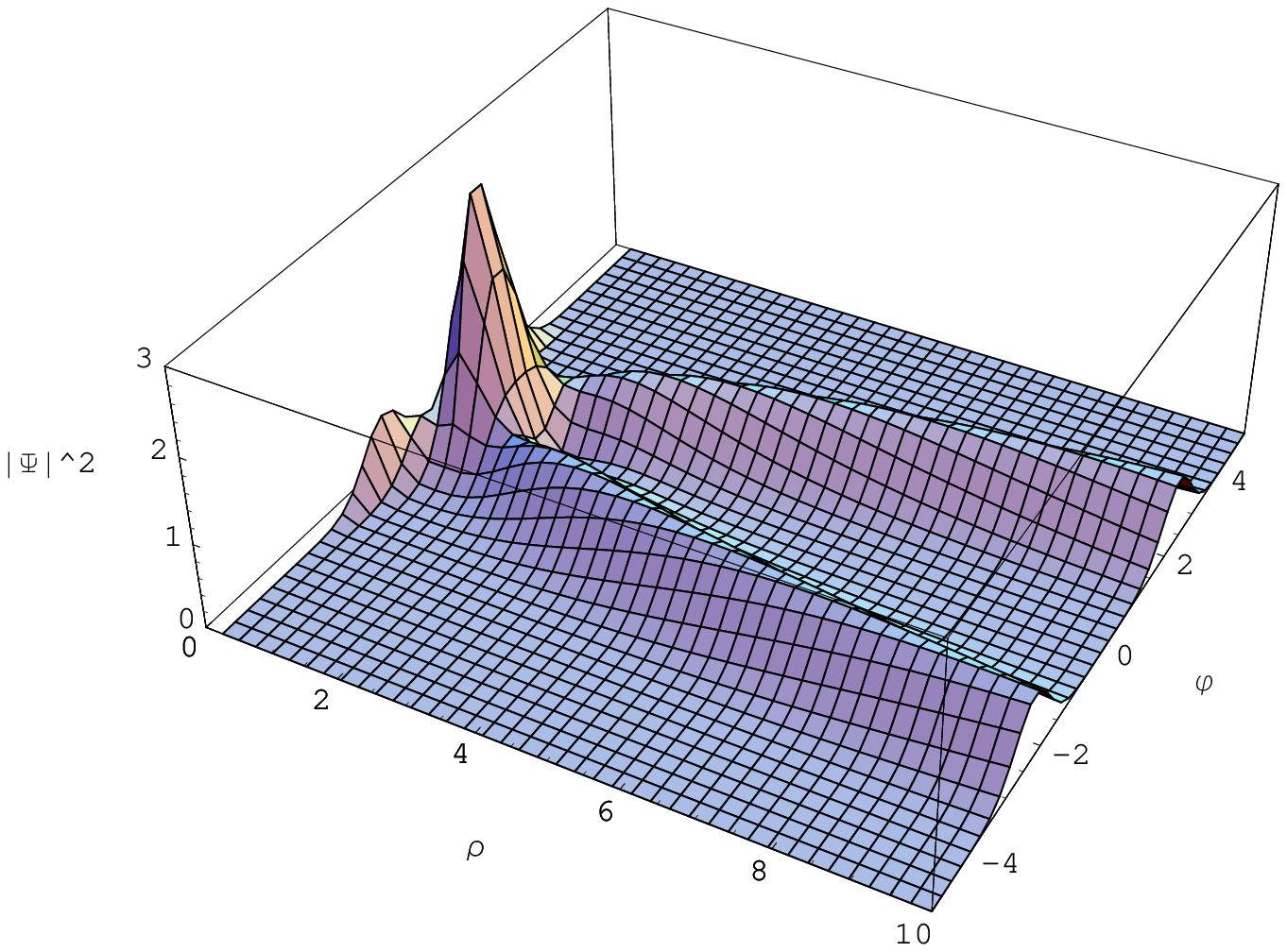,width=7cm}\hspace{5mm}
\epsfig{figure=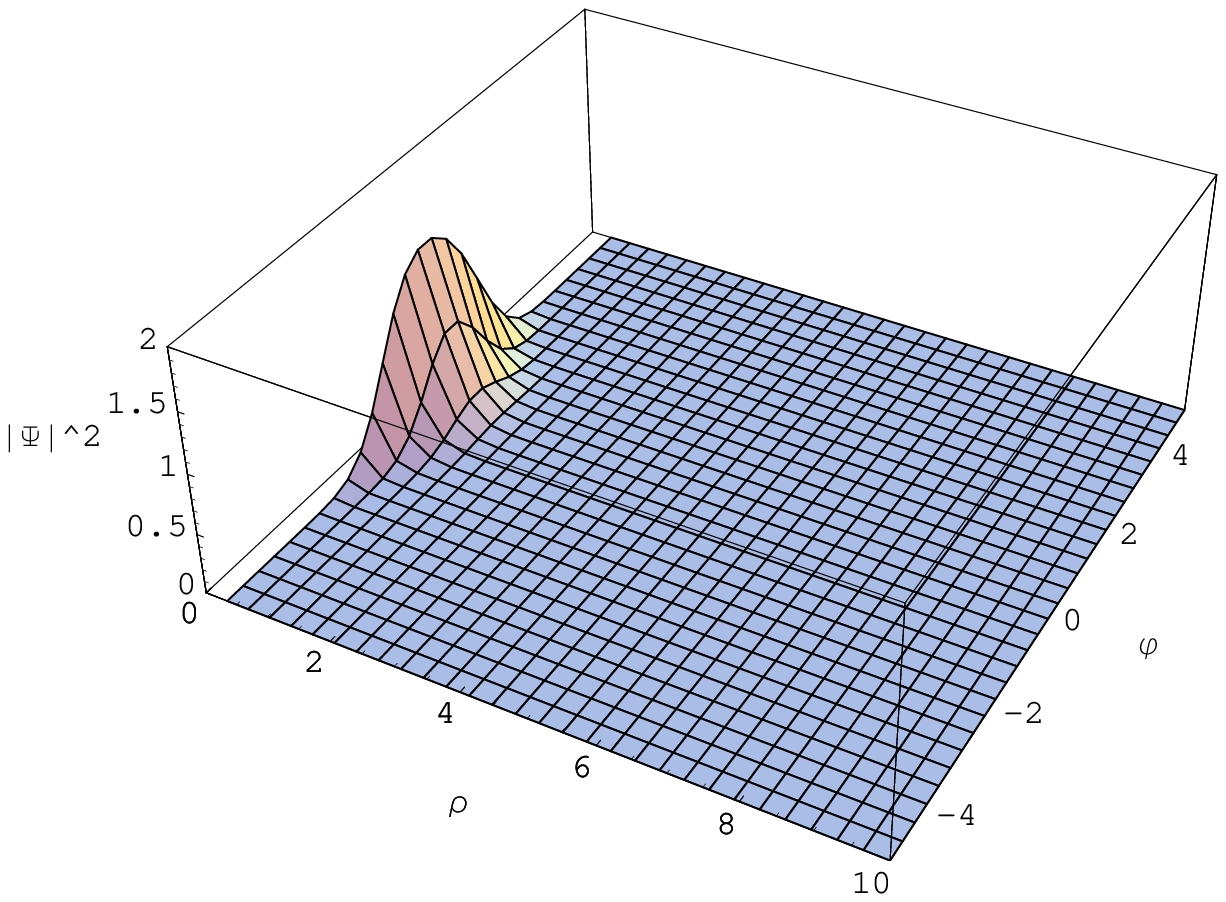,width=7cm}\vspace{5mm}
\end{center}

\hspace{1.5cm}{\footnotesize \textbf{Figure 2}: The probability
$|\Psi|^2$ in the commutative case for $k=0$ (left) and $k=1$
(right), both

\hspace{1.5cm}with the Gaussian constants $b=1=c$.}
\end{figure}
\begin{figure}
\begin{center}
\epsfig{figure=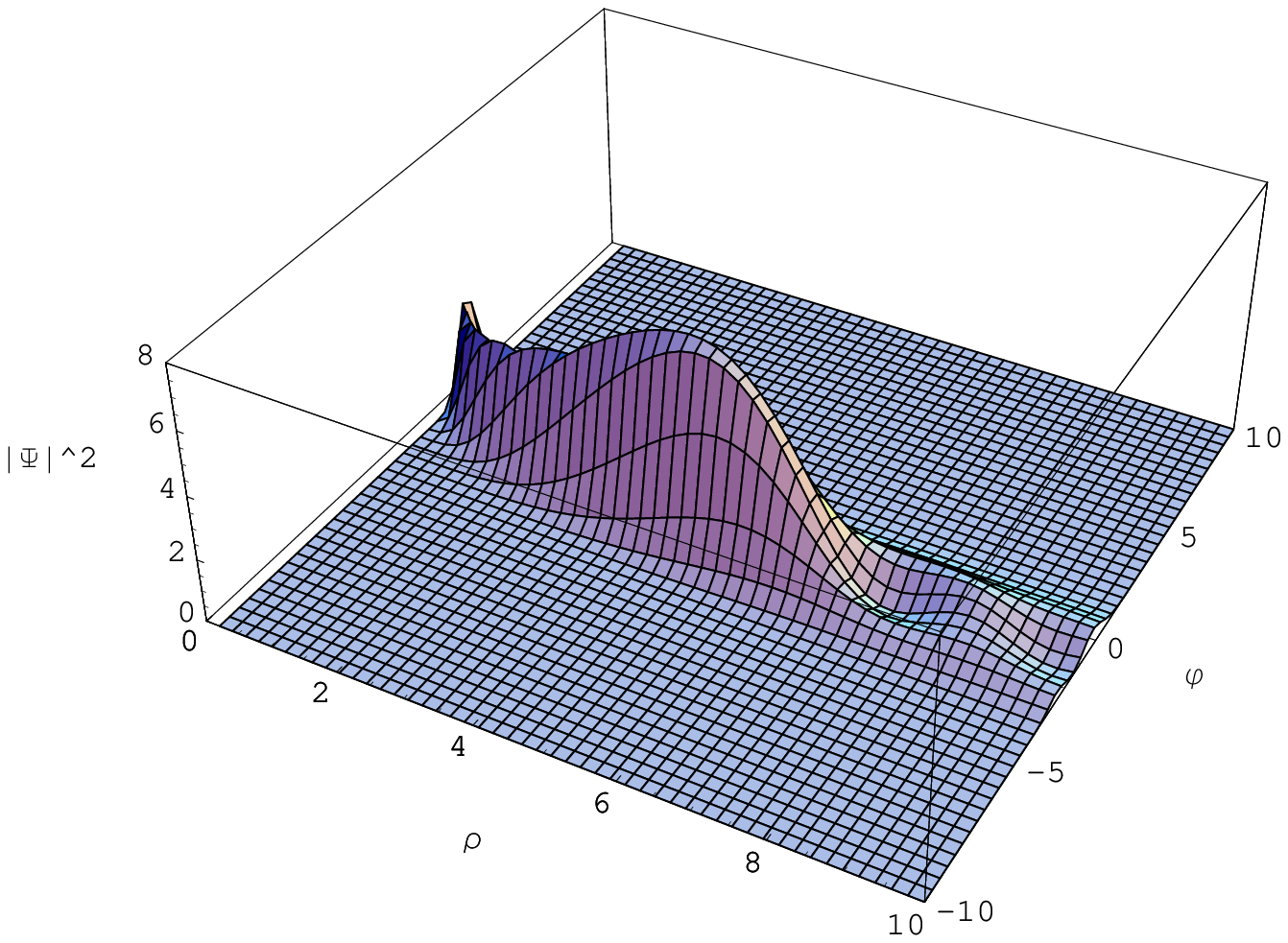,width=7cm}\hspace{5mm}
\end{center}

\hspace{0.8cm}{\footnotesize \textbf{Figure 3}: The probability
$|\Psi|^2$ in the NC case for $k=0$,\hspace{2mm}$\beta=1/10$ and
the Gaussian constants $b=1=c$.}
\end{figure}
\end{document}